\documentclass[useAMS,usenatbib,usegraphicx,usedcolumn]{mn2e}
\usepackage{multirow}
\usepackage{amsmath}
\title[Circumstellar Habitable Zones of Binary Star Systems in the Solar Neighborhood]{Circumstellar Habitable Zones of Binary Star Systems in the Solar Neighborhood}
\author[S. Eggl and  E. Pilat-Lohinger and B. Funk and N. Georgakarakos and N. Haghighipour]{S. Eggl$^{1}$\thanks{E-mail:eggl@astro.univie.ac.at (SE)}, 
E. Pilat-Lohinger$^{1}$\thanks{E-mail:elke.pilat-lohinger@univie.ac.at (EPL)},
B. Funk$^{1}$\thanks{E-mail: funk@astro.univie.ac.at (BF)}, 
N. Georgakarakos$^{2}$\thanks{E-mail: georgakarakos@hotmail.com (NG)} and
N. Haghighipour$^{3}$\thanks{E-mail: nader@ifa.hawaii.edu (NH)}\\
$^{1}$University of Vienna, IfA, T\"urkenschanzstr. 17, 1180 Vienna, Austria\\
$^{2}$Higher Technological Educational Institute of Serres, Terma Magnesias, Serres 62124, Greece\\
$^{3}$ Institute for Astronomy and NASA Astrobiology Institute, 2680 Woodlawn Drive, Honolulu, HI 96822, USA}

\begin{document}

\date{Accepted: 2012 October 19. - . Received: 2012 October 19; in original form 2012 September 16;}

\pagerange{\pageref{firstpage}--\pageref{lastpage}} \pubyear{2012}

\maketitle

\label{firstpage}

\begin{abstract}

Binary and multiple systems constitute more than half of the total stellar population in the Solar neighborhood \citep{kiseleva-eggleton-eggleton-2001}. 
Their frequent occurrence as well as the fact that more than 70 \citep{schneider-et-al-2011} planets have already been discovered in such configurations 
- most noteably the telluric companion of $\alpha$ Cen B \citep{dumusque-et-al-2012} -  
make them interesting targets in the search for habitable worlds. Recent studies \citep{eggl-et-al-2012, forgan-2012} have shown, that despite the variations in 
gravitational and radiative environment, there are indeed circumstellar regions where planets can stay within habitable insolation limits on 
secular dynamical timescales. 
In this article we provide habitable zones for 
19 near S-Type binary systems from the Hipparchos and WDS catalogues with semimajor axes between 1 and 100 AU. Hereby, we
accounted for the combined dynamical and radiative influence of the second star on the Earth-like planet. 
Out of the 19 systems presented, 17 offer dynamically stable habitable zones around at least one component.
The 17 potentially habitable systems contain 5 F, 3 G, 7 K and 16 M class stars.
As their proximity to the Solar System ($d<31$ pc) makes the selected binary stars exquisite targets for observational campaigns, 
we offer estimates on radial velocity, astrometric and transit signatures produced by habitable Earth-like 
planets in eccentric circumstellar orbits.

\end{abstract}


\section{Introduction}
The discovery and confirmation of terrestrial bodies orbiting other stars (e.g. \citet{dumusque-et-al-2012, borucki-et-al-2012,kepler2-2011,corot7b-2009}) has  
generated enormous public as well as scientific interest. It has shown that after a mere two decades of exoplanetary research 
finding potentially habitable worlds around other stars seems to be almost within our grasp. 
Close-by stars and stellar systems are thereby premium targets, as they tend to offer reasonable signal to noise ratios (SNRs) for photometry 
and radial velocity as well as comparatively large astrometric amplitudes \citep{beauge-et-al-2007,neat-2011,guedes-et-al-2008,eggl-et-al-2012b}.   
As more than half of the stars in the Solar neighborhood are members of binary or multiple systems \citep{kiseleva-eggleton-eggleton-2001}, 
it is not surprising that more than 70 planets in or around binary-stars have been discovered 
\citep{schneider-et-al-2011} despite the current observational focus on single star systems.
Even-though NASA's \textit{Kepler} mission has been quite successful in finding circumbinary planets 
(e.g. \citet{welsh-et-al-2012,orosz-et-al-2012,doyle-et-al-2011})
 we will focus on binary-star systems with potential \textit{circumstellar} habitable zones (HZs) in this study. 
In fact, most of the planets discovered in double stars are in these so-called S-Type configurations \citep{rabl-dvorak-1988, roell-et-al-2012},
where the planet orbits one star only. 
The telluric companion of $\alpha$ Cen B is such an example \citep{dumusque-et-al-2012}.

An interesting question in this regard is doubtlessly: Can S-Type binary stars harbor habitable worlds?
Already \citet{huang-1960} and \citet{harrington-1977} and more recently \citet{forgan-2012} investigated the effects such configurations have on the insolation hypothetical planets would receive. 
\citet{eggl-et-al-2012} (in the following referred to as EG12) were able to derive analytic expressions to find HZs in binary star systems 
unifying dynamical and radiative balance models for 
%practically\footnote{Planetary orbits are assumed to have negligible eccentricity 
%at the end of the gas disc dominated formation phase.} any   
S-Type binary star - planet systems.
While the exact manner in which planets form in tight binary star systems is still hotly debated
in astrophysical literature - see for instance \citet{mueller-kley-2012, batygin-et-al-2011, paardekooper-leinhardt-2010, thebault-2011} and references therein,
the discovery of \mbox{ $\alpha$ Cen B b} has made the existence of terrestrial planets in S-Type binary star systems an observational fact.
Opinions still differ on whether it is theoretically possible that planets in $\alpha$ Centauri's HZs can form on stable orbits.
Even though classical N-body simulations with best case accretion scenarios seem to be able to produce terrestrial planets 
near the HZs of the $\alpha$ Centauri system \citep{quintana-lissauer-2010, guedes-et-al-2008}, \citet{thebault-et-al-2009,thebault-et-al-2008}
concluded that even when gas drag is included the encounter velocities between kilometer sized planetesimals would lead to erosive collisions, thus
making constant accretion unlikely. However, in their model they did not include a self-consistent evolution of the
gas disc, nor did they consider planetesimal self gravitation or re-accretion of collisional debris. 
\citet{paardekooper-leinhardt-2010} used a self consistent disc model with planetesimals. They were able produced accretion friendly scenarios
when the collision frequency was sufficiently high to prevent orbital dephasing.  
Other possible solutions to the problem of high encounter velocities range from including  
planetesimal and embryo migration \citet{payne-et-al-2009} over mild inclination of planetesimal discs with respect to the binary's orbit \citep{xie-et-al-2010} to 
more realistic radiative modeling of the system's gaseous disc \citep{mueller-kley-2012}.

% A related issue is the role of mutual inclination between the disc and the binary's orbit respectively.
% \citet{xie-zhou-2009} found that a small inclination of the disc with respect to the binary's orbit can favor 
% accretion, which has been disputed by \citet{thebault-2011}.  
% 
% mueller-kley-2012
% ,quintana-lissauer-2011,thebault-et-al-2009,thebault-2008,guedes-et-al-2008}.
% accretion form planetesimals to planetary embryos. 
% 
% Especially the role of disc and planetesimal inclination with regard to the binary's orbit is controversial. xie .. claim that
% mutual inclination will improve the chances of , thebault low inclination no, fragner 3D hydro simulations 
% without self gravity kozai no.
%  mueller kley in favor more realistic treatment of radiative processes  
% condtions morbidelli tsiganis self gravitating planetesimal discs can stop kozai
% 
% a lot of open questions remain. 
% 
% but

\citet{eggl-et-al-2012b} show that even if additional Earth-like planets in $\alpha$ Centauri do exist, 
it is not an easy task to find them given current observational limitations
in radial velocity resolution. The RV signal semi amplitude of $\alpha$ Cen B b was near the current edge of feasibility with \mbox{$\delta$~RV $\simeq 50$cm/s}
, whereas accuracies lower than $10$ cm/s would be necessary to discover telluric planets in $\alpha$ Cen B's HZs.
Astrometry is not much more helpful in this case, as the necessary astrometric amplitudes to detect habitable worlds in
$\alpha$ Centauri will  only be available near the end of the GAIA mission's lifetime \citep{hestroffer-et-al-2010}. 

In this study we tackle the question whether there are S-Type systems in the Solar neighborhood that might make for easier targets. 
For this purpose we select 19 S-Type binary systems from "The Washington Visual Double Star Catalog" (WDC) \citep{mason-et-al-2012} with well determined stellar parameters that lie 
within a distance of 31 pc from the Solar System, and calculate HZs for each stellar component using the analytic method presented in \citet{eggl-et-al-2012}.
We provide estimates on the RV and astrometric (AM) root mean square (rms) signal strengths expected for an Earth-like planet orbiting at the borders 
of a system's HZs. Furthermore, we present likely transit depths for potentially transiting habitable planets in co-planar S-Type double star systems.

This article is structured as follows:
First, we will discuss the selection criteria for the 19 systems investigated (section \ref{sec:sel}). 
After a brief summary of the main factors that determine habitability for terrestrial planets in binary star systems (section \ref{sec:hz}),
the issue of dynamical stability of planets in such configurations is addressed in section \ref{sec:stab}.
Our results - tables with HZ borders and signal strength estimates - are presented and discussed in sections \ref{sec:res} and \ref{sec:trans}.
Current problems in modeling tidal locking of planets in binary systems are mentioned in section \ref{sec:tides}.
A summary (section \ref{sec:sum}) concludes this study.
 
\section{Selection of binary star systems} \label{sec:sel}
We preselected all detached binaries with semimajor axes $1<a_b<100$ AU using the stellar orbital parameters provided in the WDC,
in order to find suitable S-Type systems in the Solar neighborhood where Earth-like planets in HZs could be detectable. 
Hereby, we only considered systems within a distance of $d<31$ pc from our Solar System as determined by the Hipparchos mission \citep{vanleeuwen-2007}. 
Together with the prerequisite that the binaries' orbital elements had to be available, 
the afore mentioned restrictions reduced the number of admissible double star systems to 313.
Furthermore, only double star configurations with known spectal types of both components were used. 
Peculiar spectra that might have been classified incorrectly by Hipparchos like HIP 17544 \& 73695 were
also excluded, narrowing the set of candidate systems from 313 to 35. 
The ultimate selection criterion consisted of calculating the binaries' periods using bolometric luminosity derived masses 
together with the semimajor axes given in WDS and comparing them to the observed binary periods. 
The stellar bolometric luminosities and masses required for this purpose were derived as follows:
With the distances available through Hipparchos data the systems' absolute visual magnitudes could be calculated. 
In order to assess the bolometric luminosities of the binary sample, we performed bolometric corrections (BC) 
of the absolute visual magnitudes using the polynomial fits by \citet{flower-1996}.
The required effective temperatures were estimated via spectral type and luminosity using the \mbox{ATLAS9} catalog of stellar model atmospheres \citep{castelli-kurucz-2004}.
We then calculated the binaries' periods using the masses derived via the mass-luminosity relations given in \citet{salaris-cassisi-2005}.
Only those systems who's derived periods did not deviate more than 11\% from the observed periods were selected for the final sample.
Stellar and orbital parameters for the final set of 19 S-Type binary systems are presented in Table \ref{tab1}.

In the next section we will briefly discuss the main points on how to determine HZs for these S-Type binary systems.
\begin{table*}
 \centering
 \begin{minipage}{160mm}
  \caption{Orbital and stellar parameters of the 19 investigated binary star systems. 
The values of parameters printed in bold letters are taken from \citep{mason-et-al-2012,vanleeuwen-2007}, the others were derived as described in section \ref{sec:sel}.
The binary's eccentricity and semimajor axis are denoted by $a_b$ and $e_b$; $I$ is the system's inclination to the plane of the sky. 
A binary components' masses are symbolized by $M_A$ and $M_B$, their respective luminosities by $L_A$ and $L_B$ and their effective temperatures are denoted $T_{eff A}$ and  $T_{eff B}$.
Stellar classifications are given in the columns headed "class A" and "class B". \label{tab1}}
  \begin{tabular}{@{}r|rrrrccrrccll@{}}
  \hline
\textbf{HIP ID} & \boldmath{$a_b$}  & \boldmath{$e_b$} & \boldmath{$I$} & \boldmath{$d$} & $M_A$ & $M_B$ & $L_A$ &
 $L_B$  & $T_{eff A}$  & $T_{eff B}$ & \textbf{class A} & \textbf{class B} \\ \hline 
14669      &           9.0 &          0.14 &          96.8 &          15.8 &          0.56 &          0.39 &         0.096 &         0.026 &     3580 &     3370 & M2         & M4         \\
30920      &           4.3 &          0.37 &          51.8 &           4.1 &          0.22 &          0.08 &         0.007 &         0.001 &     3370 &     3145 & M4V        & M5.5V      \\
31711      &          42.7 &          0.34 &          93.9 &          21.3 &          1.03 &          0.57 &         1.137 &         0.109 &     5860 &     4060 & G2V        & K7Ve       \\
44248      &          10.4 &          0.15 &         131.4 &          16.1 &          1.44 &          0.89 &         4.285 &         0.638 &     6740 &     5250 & F3V        & K0V        \\
45343      &          97.2 &          0.28 &          21.0 &           5.8 &          0.52 &          0.51 &         0.073 &         0.067 &     3850 &     3850 & M0V        & M0V        \\
51986      &           9.9 &          0.75 &         129.1 &          26.8 &          1.88 &          1.29 &        12.535 &         2.790 &     6710 &     6740 & F4IV       & F3         \\
58001      &          11.7 &          0.30 &          51.0 &          25.5 &          2.94 &          0.79 &        65.255 &         0.397 &     9520 &     4780 & A0Ve       & K2V        \\
64241      &          11.8 &          0.50 &          90.1 &          17.8 &          1.30 &          1.12 &         2.887 &         1.553 &     6440 &     6360 & F5V        & F6V        \\
64797      &          89.2 &          0.12 &          93.4 &          11.1 &          0.73 &          0.52 &         0.277 &         0.072 &     5015 &     3715 & K1V        & M1V        \\
66492      &          46.9 &          0.61 &          36.3 &          22.0 &          0.59 &          0.48 &         0.121 &         0.054 &     3782 &     3647 & M0.5       & M1.5       \\
67422      &          32.7 &          0.45 &          47.4 &          13.4 &          0.72 &          0.65 &         0.273 &         0.174 &     4560 &     4205 & K4V        & K6V        \\
84425      &           7.7 &          0.49 &         115.2 &          30.6 &          1.23 &          0.86 &         2.267 &         0.556 &     6280 &     5860 & F7V        & G2V        \\
84720      &          91.6 &          0.78 &          35.6 &           8.8 &          0.79 &          0.50 &         0.393 &         0.062 &     5570 &     3850 & G8V        & M0V        \\
87895      &           2.4 &          0.41 &          68.0 &          28.2 &          1.19 &          0.90 &         2.031 &         0.648 &     5860 &     4780 & G2V        & K2V        \\
93825      &          32.7 &          0.32 &         149.6 &          17.3 &          1.27 &          1.25 &         2.570 &         2.432 &     6200 &     6200 & F8V        & F8V        \\
101916     &          15.7 &          0.80 &         107.0 &          30.1 &          1.61 &          0.37 &         6.794 &         0.023 &     5745 &     4420 & G1IV       & K2IV       \\
106972     &           5.3 &          0.29 &          69.4 &          24.5 &          0.57 &          0.43 &         0.105 &         0.033 &     3370 &     3370 & M2         & M4         \\
114922     &           6.7 &          0.44 &         117.1 &          30.8 &          0.49 &          0.52 &         0.059 &         0.073 &     3715 &     3580 & M1         & M2         \\
116132     &          42.5 &          0.20 &         123.5 &           6.2 &          0.38 &          0.20 &         0.025 &         0.006 &     3370 &     3305 & M4         & M5         \\
\hline
&  [AU] & & [deg] &[pc] &[$M_\odot$] & [$M_\odot$] & [$L_\odot$] & [$L_\odot$] & [K] & [K] & &\\
\hline
\end{tabular}
\end{minipage}
\end{table*}

\section{Habitability of Earth-like planets in S-Type binary star systems} \label{sec:hz}
The most pronounced difference between determining classical HZs and HZs for Earth-like planets in binary star systems 
lies in the assumption that planetary orbits are basically circular.
In fact, the well known borders defined by \citet{kasting-et-al-1993} are built on the premises that planetary insolation will change only
on stellar evolutionary timescales. Thus, the planet is thought to remain more or less at the same distance from its host star on a circular orbit.
This assumption is implicitly made in almost all recent works, e.g. \citet{kane-gelino-2012, pierrehumbert-gaidos-2011, kaltenegger-sasselov-2011}.
However, in three body systems, such as the planet - binary star configurations we are investigating, gravitational interactions
will alter the planetary orbit. 

Perturbation theory of hierarchical triples predicts, that the orbit of the inner pair - in our case host-star and planet - 
will experience significant alterations in its eccentricity, whereas its semimajor axis remains almost constant 
\citep{marchal-1990,georgakarakos-2002,georgakarakos-2003}.
For nearly equiplanar systems, the influence of planetary inclination and ascending node to the overall dynamics can be considered small, they will be neglected in what follows. 
Even-though there may be short periodic variations, some important changes in a planet's orbit happen also on secular timescales.
Secular periods are usually much larger than the planet's orbital period. 
However, they are a lot smaller than stellar evolutionary timescales for detached binary systems with semimajor axes $a_b<100$ AU. 
It is thus necessary to include the effects of changing planetary orbits in our estimates regarding HZs within binary star environments.
In their work, EG12 confirmed that variations in the planet's orbit are even more important for changes in its insolation than the additional radiation form the second star!
The only exceptions to this rule are systems where the second star is much more luminous than the planet's host-star ($L_B/L_A>4$, where binary component $A$ 
is the planet's host-star in this case).
Therefore, a planet's eccentricity is a dominating factor in determining habitability.
Yet, how eccentric can a planetary orbit become, in order to still allow for habitability?
 
\citet{williams-pollard-2002} concluded that 
an Earth-like atmosphere together with surface oceans can buffer the harsh changes between high insolation at periastron and long cold phases near apoastron up to eccentricities of $e_p\approx0.7$, as long
as the average insolation is comparable with the current insolation of the Earth. 
Although planetary eccentricities of such magnitude are usually not reached in close S-Type setups (EG12), the region where the planet remains within 
classical insolation boundaries is still strongly impacted. 
In order to distinguish orbital zones that are only habitable "on average" and zones where the planet will never exceed classical insolation limits,
EG12 introduced three types of HZs for binary star systems:   

\begin{description}
\item[\textbf{Permanently Habitable Zone} (\textbf{PHZ}): ]
The PHZ is the region where a planet stays within habitable insolation limits for all times, despite
the changes its orbit experiences due to gravitational interactions with the secondary.  
For this study, we have chosen the classical runaway/maximum greenhouse insolation limits as defined by \citet{kasting-et-al-1993} and \citet{underwood-et-al-2003} (\textbf{KHZ})
\item[\textbf{Extended Habitable Zone} (\textbf{EHZ}): ]
The binary-planet configuration is still considered to be habitable when 
most of its orbit remains within the HZ boundaries. This is true if the average received insolation plus one standard deviation
does not put the planet beyond KHZ insolation limits.
\item[\textbf{Averaged Habitable Zone} (\textbf{AHZ}): ]
Even an elevated planetary eccentricity ($e<0.7$) may not be prohibitive for habitability since the atmosphere
acts as a buffer \citep{williams-pollard-2002}, if the time averaged insolation stays within habitable limits. 
The AHZ represents such regions.
\end{description}
For details on the definition and calculation of PHZ, EHZ \& AHZ we refer the reader to EG12.
We use the interpolation formulae given in \citet{underwood-et-al-2003} to calculate effective insolation values for the selected stellar types. 
After a brief discussion concerning aspects of dynamical stability, 
the application of the proposed classification scheme to the 19 selected binary star systems will be presented in the next section.

\section{Dynamical Stability of Circumstellar Planets in Binary Stars}\label{sec:stab}
As was briefly mentioned during the introduction, there are many open questions regarding the formation of planets in double star environments \citep{thebault-2011}. However,
once formed a planet can survive in the dynamically stable region around
one of the binary components - a fact proven by observed planets in S-Type binary configurations \citep{dumusque-et-al-2012,roell-et-al-2012,giuppone-et-al-2012}. 
If the necessary dynamical prerequisites are fulfilled, even both stars can harbor planets. Generalized dynamical investigations such as
\citet{holman-wiegert-1999}, \citet{pilat-lohinger-dvorak-2002}, semi-analytical \citep{pichardo-et-al-2005} or analytical approaches \citep{szebehely-mckenzie-1977,eggleton-1983} 
can be used to determine regions where a test-planet can remain on a stable orbit on secular dynamical timescales. 
As the setup used in this work consists of a planar binary - Earth configuration, the restricted three body approach 
used in the articles mentioned above can be considered a reasonable approximation.
We will apply the numerical fit by \citet{holman-wiegert-1999} and results by \citet{pilat-lohinger-dvorak-2002} to find critical semimajor axis for circumstellar motion.

\section{Results}\label{sec:res}
\begin{figure}
\begin{tabular}{cc}
\includegraphics[angle=180, scale=0.48]{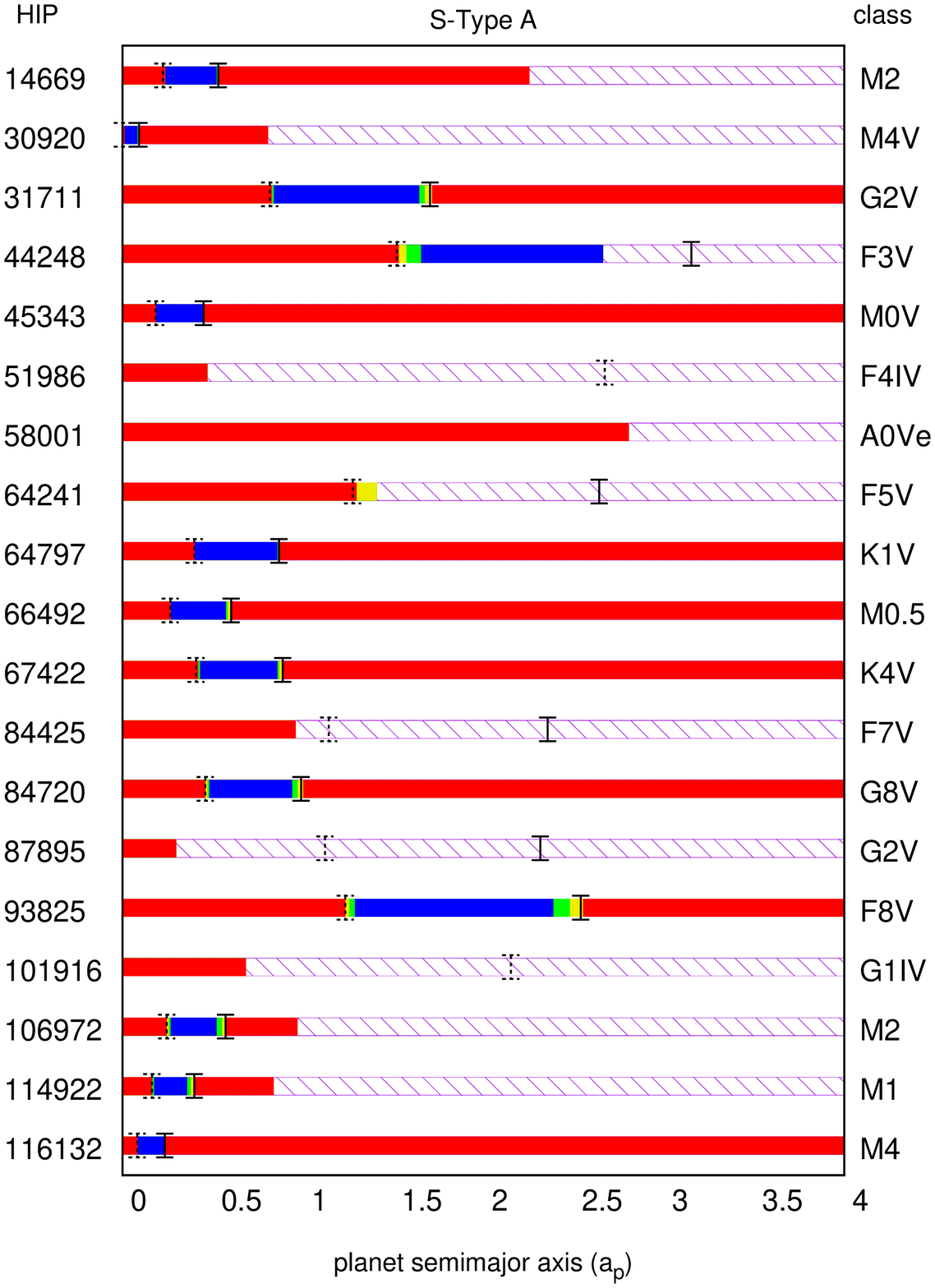}
\includegraphics[angle=180, scale=0.48]{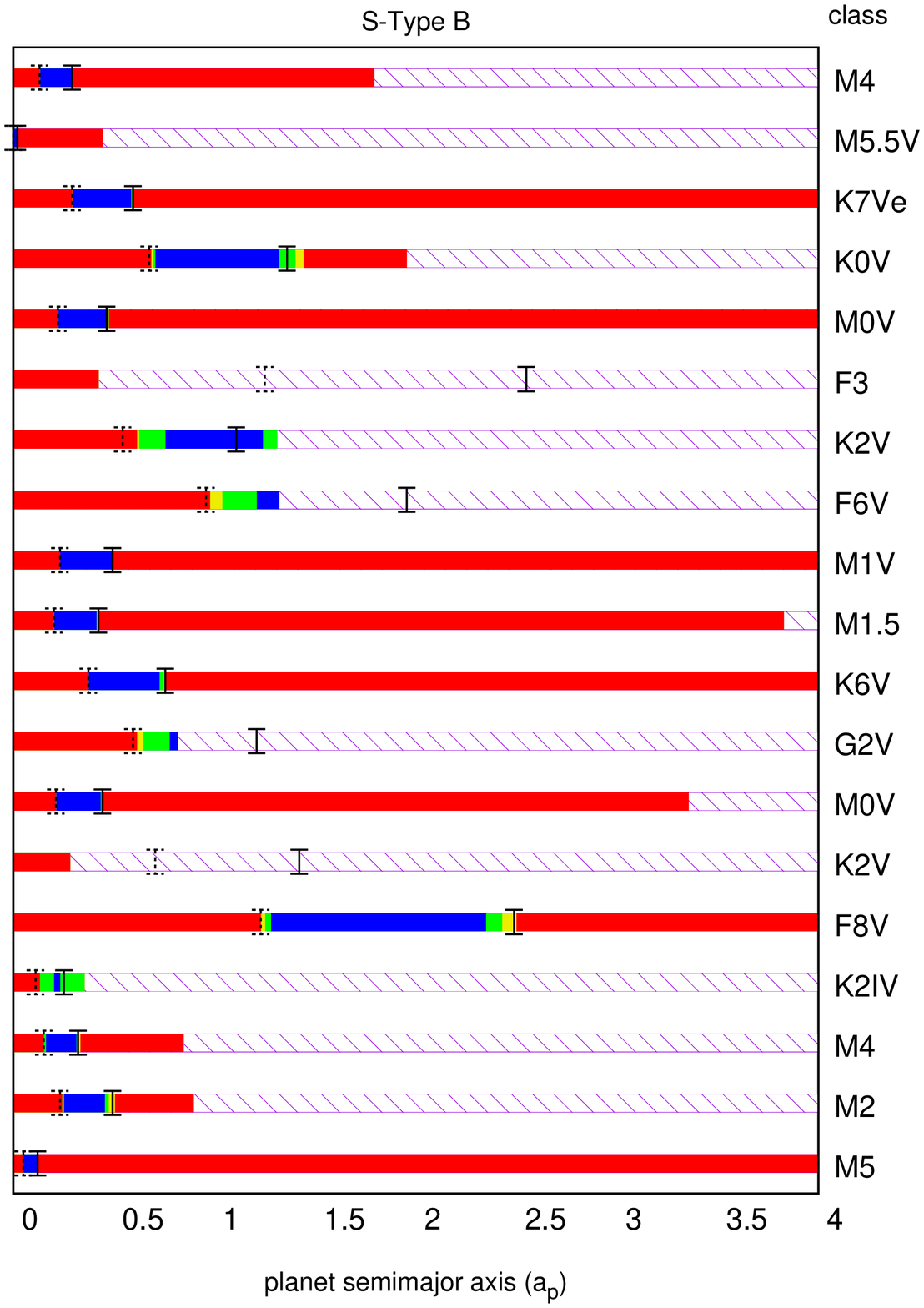}  
\end{tabular}
\caption{Habitable zones of 19 S-Type binary star systems in the Solar neighborhood are shown. The light gray regions (blue online) denote
zones of permanent habitability (PHZ), medium gray (green online) Extended (EHZ) and dark gray (yellow online) Averaged Habitable Zones (AHZ), see
section \ref{sec:hz}.
Black (red online) are regions where the planet either receives too much, or too little radiation to keep atmospheric temperatures stable.
The striped areas are zones of dynamical instability \citep{holman-wiegert-1999}. \textit{Left:} HZs around the system's primary star are shown (S-Type A), 
\textit{right:} habitability of regions around the secondary are investigated (\mbox{S-Type B}) \citep{whitmire-et-al-1998}.
The dashed 'I' symbols indicate the inner, the full symbols the outer border of the classical HZ as defined by \citet{kasting-et-al-1993} \& \citet{underwood-et-al-2003}.
In most cases, the AHZ and the classical HZ coincide well as was pointed out in \citep{eggl-et-al-2012}, except for the
systems HIP 58001 and 101916 where the considerable luminosity of the brighter companions shift the HZs of the \mbox{S-Type B} configurations to larger planetary semimajor axes. 
Evidently, 17 out of the 19 investigated systems allow for dynamically stable terrestrial planets within HZs around at least one of its binary's components.  \label{fig1}}
\end{figure}

The different HZs discussed in section \ref{sec:hz} are presented for a fictitious Earth-like planet in each of the selected double star systems (Fig. \ref{fig1}). 
The region of instability (striped) is also marked. The left graph of Fig.~\ref{fig1} represents HZs around the primary (S-Type A), 
and the right graph shows HZs around the secondary (S-Type B) \citep{whitmire-et-al-1998}.
Black (red online) denotes regions which are non-habitable due to excessive or insufficient insolation, dark gray (yellow online), medium gray (green online) 
and light gray (blue online) represent the AHZ, EHZ and PHZ respectively.
Dashed and full 'I' symbols give the inner and outer borders of the classical HZ as defined by \citet{kasting-et-al-1993} and \citet{underwood-et-al-2003} (KHZ).
\citet{eggl-et-al-2012} found a good correspondence between the KHZ and the AHZ, which is also mirrored in the results
at hand. Exceptions are the systems HIP 58001 \& 101916 where the more luminous companion shifts the HZs of the less luminous one considerably.
Out of the 19 selected systems, 17 permit Earth-like planets in HZs on dynamically stable orbits around at least one stellar component. 
In total, the 17 habitable systems feature 16 M, 7 K, 3 G and 5 F class stars. Even if the all F and M class stars were to be excluded from the list of 
hosts for HZ - either because of their comparatively short lifespans \citep{kasting-et-al-1993} or tidal and radiative effects (see section \ref{sec:tides}) - 
more than 26\% of the stars in this sample would be capable of sustaining habitable planets on secular dynamical timescales.
If the stars' mass loss via stellar winds is negligible, and no cataclysmic events occur \citep{veras-tout-2012}, habitability might be given even for stellar evolutionary timescales.  

A detailed listing of HZ-borders as well as expected radial velocity (RV) and astrometric (AM) signal strengths produced by a terrestrial planet in the selected systems is presented in Tables \ref{tab2} - \ref{tab4}.
Maximum and root mean square (rms)\footnote{In this case rms values have not only been time averaged, but they were also averaged over the planet's argument of pericenter.} 
signal strengths have been calculated following \citet{eggl-et-al-2012b}. 
The corresponding equations are repeated in appendix \ref{ap:amp}  for the reader's convenience.  
Comparing AM and RV signal strengths one can see that - current observational equipment assumed - RV seems to stand a better chance to find Earth-like planets in 
HZs of nearby double stars. 
% \citet{guedes-et-al-2008} claim, that
% even RV signal to noise ratios (SNR) down to 0.1 may not be prohibitive for finding terrestrial planets in binary stars like $\alpha$ Centauri.
With the discovery of $\alpha$ Cen B b the currently feasible RV resolution is approximately $50$ cm/s.   
For the detection of habitable planets in the  $\alpha$ Centauri system, however, semi-amplitudes around $10$ cm/s would be required \citep{eggl-et-al-2012b}. 
Possible candidate systems such as HIP 14699, 30920, 106972, 114922 or 116132  would offer better conditions for finding
habitable Earth analogues via RV than \mbox{$\alpha$ Centauri} does. 

As 9 out of the 17 potentially habitable systems feature M-stars, it is worth mentioning that determining the effective insolation a terrestrial planet receives
might not be enough to claim habitability. In fact, \citet{lammer-et-al-2011} are convinced that the potentially elevated level of X-ray and extreme UV radiation 
in M-stars might lead to a different atmospheric evolution of an Earth-like planet in an M-star's HZ, thus  preventing the existence of life as we know it.
Ultimately direct observation of the interaction between stellar and planetary atmospheres will be necessary to determine to which degree
planets can remain habitable in the vicinity of M-type stars. The proposed transit spectroscopy mission ECHO \citep{tinetti-et-al-2012} 
can be a step in this direction, although currently only super-Earths down to $1.5\; r_\oplus$ around K-F stars are planned to be observed.
With RV signal amplitudes of $\approx 5-12$ cm/s for potentially habitable planets in systems containing Sun-like G stars (HIP 31711 \& 84425), 
our estimates are comparable to those for $\alpha$ Centauri presented in \citet{eggl-et-al-2012b} and \citet{guedes-et-al-2008}.  
Detecting planets around Sun-like stars would therefore require a considerable amount of dedicated observation time \citep{dumusque-et-al-2012,guedes-et-al-2008}.

The AM amplitudes determined for the 19 systems at hand are well below $1\;\mu as$. This will put the systems in consideration 
even beyond the reach of ESA's GAIA mission \citep{hestroffer-et-al-2010}. 
However, recently \citet{neat-2011} proposed the Nearby Earth Astrometric Telescope (NEAT)
which would be capable of resolving astrometric motion down to $0.05\;\mu as$ at a one $\sigma$ accuracy level. This instrument would be able to identify 
habitable planets in most of the presented binary star systems. Such a mission would indeed be valuable, since 
AM does not only favor planet detection in binary configurations with Sun-like components - their
HZs are further away form their host stars thus producing larger AM amplitudes - it would more importantly grant observational access to all the planet's orbital parameters.
Especially mutual inclinations are of interest in this case, as they could provide answers to many important problems regarding planet-formation as well as
migration in binary star systems \citep{batygin-et-al-2011, thebault-2011, wu-murray-2003}.

\section{Potentially Transiting Systems}\label{sec:trans}
With an inclination  of $I\approx90^\circ$ with respect to the plane of the sky the systems HIP 14669, 31711, 64241 \& 64797 could harbor transiting planets that
still would be compatible with our assumptions of a planar binary planet configuration.  
% Where a planet could be transiting and is still compatible with the assumption of a planar system ($i_p-i_b\approx0$).
Assuming non grazing transits, i.e. transits where less than the full planetary disc obscures the stellar surface during transit, 
and neglecting entry as well as limb darkening effects, we can estimate the 
relative transit depth ($TD$) that the planet will cause in its host-star's photometric signal:
\begin{equation}
TD \simeq \frac{R_p^2}{R_\star^2}
\end{equation}
Hereby, $R_p$ and $R_\star$ denote the planetary and stellar radii respectively.
Table \ref{tab5} shows the relative transit depths for Earth-like planets in systems allowing for transits while still being close to planar.  
Even-though some stellar components are on the verge of being too bright to be observed by \textit{Kepler}, the 
spacecraft's current performance (combined noise level $\approx 29$ ppm, \citet{gilliland-et-al-2011}), would allow for an Earth-like planet in circumstellar HZs to be found 
in all of these systems given sufficient observation time.

\begin{table*}
 \centering
 \begin{minipage}{160mm}
  \caption{Critical semimajor axis ($a_{crit}$ [AU], col. 3) for orbital stability and borders for the HZs ([AU], cols. 5-9) as measured for the respective host stars A\&B are given for 19 binary star systems
in the Solar neighborhood. Additionally, rms radial velocity (RV [cm/s]) and astrometric (AM [$\mu as$]) signatures of terrestrial planets have been evaluated at the HZ borders.
The conditions required for a planet to be withing the Averaged (AHZ), Extended (EHZ) and Permanent (PHZ) Habitable Zone are discussed in section \ref{sec:hz}.
Dashed fields (-) represent cases where a given HZ border lies beyond the critical semimajor axis $a_{crit}$. Planets there would 
be on dynamically unstable orbits. \label{tab2}}
  \begin{tabular}{@{}c|c|c|ccccccl@{}}\hline
HIP ID & comp. & $a_{crit}$ &  inner AHZ &  inner EHZ &  inner PHZ  & outer PHZ  & outer EHZ   & outer AHZ & \\ \hline 
\multirow{10}{*}{14669     } & \multirow{5}{*}{A (M2)}  & \multirow{5}{*}{        2.287} &          0.306 &         0.308 &         0.310 &         0.590 &         0.596 &         0.604 & HZ \\
 & &  &         22.02 &         21.95 &         21.88 &         16.04 &         15.96 &         15.86 & max RV \\
 & &  &         15.39 &         15.34 &         15.29 &         11.09 &         11.03 &         10.96 & rms RV \\
 & &   &         0.107 &         0.108 &         0.109 &         0.209 &         0.211 &         0.214& max AM \\
 & &  &         0.076 &         0.076 &         0.077 &         0.146 &         0.147 &         0.149 & rms AM \\\cline{2-10}
 &  \multirow{5}{*}{B (M4)}  &  \multirow{5}{*}{        1.806}  &         0.162 &         0.162 &         0.162 &         0.316 &         0.318 &         0.320& HZ \\
 & &  &         36.00 &         36.00 &         36.00 &         25.94 &         25.86 &         25.78 & max RV \\
 & &  &         25.30 &         25.30 &         25.30 &         18.12 &         18.06 &         18.00 & rms RV \\
 & &  &         0.081 &         0.081 &         0.081 &         0.159 &         0.160 &         0.161  & max AM \\
 & &  &         0.057 &         0.057 &         0.057 &         0.112 &         0.112 &         0.113 & rms AM  \\\hline
\multirow{10}{*}{30920     } & \multirow{5}{*}{A (M4V)}  & \multirow{5}{*}{        0.865} &          0.086 &         0.086 &         0.088 &         0.162 &         0.166 &         0.168 & HZ \\
 & &  &         52.37 &         52.37 &         51.80 &         38.90 &         38.47 &         38.26 & max RV \\
 & &  &         36.24 &         36.24 &         35.83 &         26.43 &         26.11 &         25.95 & rms RV \\
 & &   &         0.291 &         0.291 &         0.298 &         0.557 &         0.571 &         0.578& max AM \\
 & &  &         0.237 &         0.237 &         0.242 &         0.445 &         0.456 &         0.462 & rms AM \\\cline{2-10}
 &  \multirow{5}{*}{B (M5.5V)}  &  \multirow{5}{*}{        0.470}  &         0.027 &         0.027 &         0.027 &         0.051 &         0.051 &         0.051& HZ \\
 & &  &        157.79 &        157.79 &        157.79 &        115.07 &        115.07 &        115.07 & max RV \\
 & &  &        110.84 &        110.84 &        110.84 &         80.36 &         80.36 &         80.36 & rms RV \\
 & &  &         0.255 &         0.255 &         0.255 &         0.487 &         0.487 &         0.487  & max AM \\
 & &  &         0.210 &         0.210 &         0.210 &         0.400 &         0.400 &         0.400 & rms AM  \\\hline
\multirow{10}{*}{31711     } & \multirow{5}{*}{A (G2V)}  & \multirow{5}{*}{        8.351} &          0.886 &         0.894 &         0.902 &         1.694 &         1.724 &         1.756 & HZ \\
 & &  &          9.63 &          9.59 &          9.55 &          7.09 &          7.03 &          6.97 & max RV \\
 & &  &          6.68 &          6.65 &          6.62 &          4.83 &          4.79 &          4.74 & rms RV \\
 & &   &         0.125 &         0.126 &         0.127 &         0.243 &         0.247 &         0.252& max AM \\
 & &  &         0.087 &         0.088 &         0.088 &         0.166 &         0.169 &         0.172 & rms AM \\\cline{2-10}
 &  \multirow{5}{*}{B (K7Ve)}  &  \multirow{5}{*}{        5.848}  &         0.316 &         0.318 &         0.320 &         0.614 &         0.618 &         0.622& HZ \\
 & &  &         21.36 &         21.29 &         21.23 &         15.42 &         15.37 &         15.32 & max RV \\
 & &  &         15.00 &         14.95 &         14.90 &         10.76 &         10.72 &         10.69 & rms RV \\
 & &  &         0.079 &         0.080 &         0.080 &         0.155 &         0.156 &         0.157  & max AM \\
 & &  &         0.056 &         0.056 &         0.056 &         0.108 &         0.109 &         0.109 & rms AM  \\\hline
\multirow{10}{*}{44248     } & \multirow{5}{*}{A (F3V)}  & \multirow{5}{*}{        2.686} &          1.581 &         1.619 &         1.697 &         2.686 &         2.686 &         2.686 & HZ \\
 & &  &          4.80 &          4.75 &          4.66 &          - &          - &          - & max RV \\
 & &  &          3.19 &          3.15 &          3.08 &          - &          - &          - & rms RV \\
 & &   &         0.221 &         0.226 &         0.238 &         - &         - &         -& max AM \\
 & &  &         0.176 &         0.181 &         0.189 &         - &         - &         - & rms AM \\\cline{2-10}
 &  \multirow{5}{*}{B (K0V)}  &  \multirow{5}{*}{        1.967}  &         0.710 &         0.718 &         0.734 &         1.340 &         1.418 &         1.456& HZ \\
 & &  &          8.76 &          8.71 &          8.62 &          6.59 &          6.44 &          6.38 & max RV \\
 & &  &          6.03 &          6.00 &          5.93 &          4.39 &          4.27 &          4.21 & rms RV \\
 & &  &         0.154 &         0.156 &         0.160 &         0.300 &         0.320 &         0.329  & max AM \\
 & &  &         0.127 &         0.129 &         0.132 &         0.241 &         0.255 &         0.262 & rms AM  \\\hline
\multirow{10}{*}{45343     } & \multirow{5}{*}{A (M0V)}  & \multirow{5}{*}{       17.932} &          0.263 &         0.263 &         0.263 &         0.515 &         0.515 &         0.517 & HZ \\
 & &  &          8.79 &          8.79 &          8.79 &          6.30 &          6.30 &          6.28 & max RV \\
 & &  &          6.20 &          6.20 &          6.20 &          4.43 &          4.43 &          4.43 & rms RV \\
 & &   &         0.265 &         0.265 &         0.265 &         0.520 &         0.520 &         0.522& max AM \\
 & &  &         0.256 &         0.256 &         0.256 &         0.501 &         0.501 &         0.503 & rms AM \\\cline{2-10}
 &  \multirow{5}{*}{B (M0V)}  &  \multirow{5}{*}{       17.698}  &         0.252 &         0.252 &         0.254 &         0.494 &         0.496 &         0.496& HZ \\
 & &  &          9.08 &          9.08 &          9.04 &          6.50 &          6.48 &          6.48 & max RV \\
 & &  &          6.41 &          6.41 &          6.38 &          4.58 &          4.57 &          4.57 & rms RV \\
 & &  &         0.259 &         0.259 &         0.261 &         0.509 &         0.511 &         0.511  & max AM \\
 & &  &         0.250 &         0.250 &         0.252 &         0.491 &         0.493 &         0.493 & rms AM  \\\hline
\multirow{2}{*}{51986     } & A (F4IV) &         0.545  & - & - & - & - & - & -  & HZ\\\cline{2-10}
 &  B (F3) &         0.448  & - & - & - & - & - & -  & HZ  \\\hline
\multirow{2}{*}{58001     } & A (A0Ve) &         2.828  & - & - & - & - & - & -  & HZ\\\cline{2-10}
 &  \multirow{5}{*}{B (K2V)}  &  \multirow{5}{*}{        1.331}  &         0.639 &         0.653 &         0.775 &         1.263 &         1.331 &         1.331& HZ \\
 & &  &         10.31 &         10.21 &          9.46 &          7.85 &          - &          - & max RV \\
 & &  &          6.98 &          6.91 &          6.34 &          4.97 &          - &          - & rms RV \\
 & &  &         0.100 &         0.102 &         0.123 &         0.210 &         - &         -  & max AM \\
 & &  &         0.080 &         0.082 &         0.097 &         0.159 &         - &         - & rms AM  \\\hline

\end{tabular}
\end{minipage}
\end{table*}

\begin{table*}
 \centering
 \begin{minipage}{160mm}
  \caption{Continuation of Table \ref{tab2}. Radial velocity (RV) amplitudes are given in [cm/s], astrometric (AM) amplitudes in [$\mu as$]. 
  The critical planetary semimajor axis $a_{crit}$ as well as the HZ borders are given in [AU].\label{tab3}}
  \begin{tabular}{@{}c|c|c|ccccccl@{}}\hline
HIP ID & comp. & $a_{crit}$ &  inner AHZ &  inner EHZ &  inner PHZ  & outer PHZ  & outer EHZ   & outer AHZ & \\ \hline 
\multirow{10}{*}{64241     } & \multirow{5}{*}{A (F5V)}  & \multirow{5}{*}{1.465} &          1.354 &         1.465 &         1.465 &         1.465 &         1.465 &         1.465 & HZ \\
 & &  &          8.19 &          - &          - &          - &          - &          - & max RV \\
 & &  &          4.82 &          - &          - &          - &          - &          - & rms RV \\
 & &   &         0.209 &         - &         - &         - &         - &         -& max AM \\
 & &  &         0.126 &         - &         - &         - &         - &         - & rms AM \\\cline{2-10}
 &  \multirow{5}{*}{B (F6V)}  &  \multirow{5}{*}{        1.339}  &         1.002 &         1.056 &         1.226 &         1.339 &         1.339 &         1.339& HZ \\
 & &  &          9.76 &          9.58 &          9.15 &          - &          - &          - & max RV \\
 & &  &          6.05 &          5.90 &          5.47 &          - &          - &          - & rms RV \\
 & &  &         0.173 &         0.184 &         0.219 &         - &         - &         -  & max AM \\
 & &  &         0.109 &         0.115 &         0.133 &         - &         - &         - & rms AM  \\\hline
\multirow{10}{*}{64797     } & \multirow{5}{*}{A (K1V)}  & \multirow{5}{*}{       23.212} &          0.472 &         0.472 &         0.474 &         0.924 &         0.926 &         0.926 & HZ \\
 & &  &         15.48 &         15.48 &         15.45 &         11.08 &         11.07 &         11.07 & max RV \\
 & &  &         10.93 &         10.93 &         10.91 &          7.81 &          7.80 &          7.80 & rms RV \\
 & &   &         0.179 &         0.179 &         0.180 &         0.351 &         0.352 &         0.352& max AM \\
 & &  &         0.126 &         0.126 &         0.127 &         0.248 &         0.248 &         0.248 & rms AM \\\cline{2-10}
 &  \multirow{5}{*}{B (M1V)}  &  \multirow{5}{*}{       18.564}  &         0.263 &         0.263 &         0.263 &         0.517 &         0.517 &         0.517& HZ \\
 & &  &         24.51 &         24.51 &         24.51 &         17.50 &         17.50 &         17.50 & max RV \\
 & &  &         17.32 &         17.32 &         17.32 &         12.35 &         12.35 &         12.35 & rms RV \\
 & &  &         0.140 &         0.140 &         0.140 &         0.275 &         0.275 &         0.275  & max AM \\
 & &  &         0.099 &         0.099 &         0.099 &         0.194 &         0.194 &         0.194 & rms AM  \\\hline
\multirow{10}{*}{66492     } & \multirow{5}{*}{A (M0.5)}  & \multirow{5}{*}{        4.289} &          0.339 &         0.341 &         0.345 &         0.645 &         0.655 &         0.667 & HZ \\
 & &  &         12.20 &         12.17 &         12.10 &          8.99 &          8.92 &          8.85 & max RV \\
 & &  &          8.48 &          8.46 &          8.41 &          6.15 &          6.10 &          6.05 & rms RV \\
 & &   &         0.081 &         0.081 &         0.082 &         0.156 &         0.158 &         0.161& max AM \\
 & &  &         0.072 &         0.072 &         0.073 &         0.137 &         0.139 &         0.142 & rms AM \\\cline{2-10}
 &  \multirow{5}{*}{B (M1.5)}  &  \multirow{5}{*}{        3.835}  &         0.227 &         0.229 &         0.231 &         0.439 &         0.443 &         0.449& HZ \\
 & &  &         16.40 &         16.33 &         16.26 &         11.92 &         11.87 &         11.80 & max RV \\
 & &  &         11.46 &         11.41 &         11.36 &          8.24 &          8.21 &          8.15 & rms RV \\
 & &  &         0.066 &         0.066 &         0.067 &         0.129 &         0.130 &         0.132  & max AM \\
 & &  &         0.059 &         0.060 &         0.060 &         0.114 &         0.115 &         0.117 & rms AM  \\\hline
\multirow{10}{*}{67422     } & \multirow{5}{*}{A (K4V)}  & \multirow{5}{*}{        4.503} &          0.486 &         0.490 &         0.496 &         0.916 &         0.934 &         0.952 & HZ \\
 & &  &         11.47 &         11.43 &         11.36 &          8.51 &          8.43 &          8.36 & max RV \\
 & &  &          7.95 &          7.92 &          7.87 &          5.79 &          5.74 &          5.68 & rms RV \\
 & &   &         0.155 &         0.157 &         0.159 &         0.298 &         0.304 &         0.310& max AM \\
 & &  &         0.130 &         0.131 &         0.133 &         0.245 &         0.250 &         0.255 & rms AM \\\cline{2-10}
 &  \multirow{5}{*}{B (K6V)}  &  \multirow{5}{*}{        4.212}  &         0.398 &         0.400 &         0.404 &         0.754 &         0.766 &         0.780& HZ \\
 & &  &         13.37 &         13.34 &         13.27 &          9.85 &          9.78 &          9.70 & max RV \\
 & &  &          9.30 &          9.27 &          9.23 &          6.75 &          6.70 &          6.64 & rms RV \\
 & &  &         0.142 &         0.143 &         0.144 &         0.273 &         0.277 &         0.282  & max AM \\
 & &  &         0.119 &         0.120 &         0.121 &         0.226 &         0.229 &         0.234 & rms AM  \\\hline
\multirow{2}{*}{84425     } & A (F7V) &         1.024  & - & - & - & - & - & -  & HZ\\\cline{2-10}
 &  \multirow{5}{*}{B (G2V)}  &  \multirow{5}{*}{        0.835}  &         0.635 &         0.667 &         0.797 &         0.835 &         0.835 &         0.835& HZ \\
 & &  &         12.54 &         12.32 &         11.67 &          - &          - &          - & max RV \\
 & &  &          7.82 &          7.63 &          6.98 &          - &          - &          - & rms RV \\
 & &  &         0.082 &         0.087 &         0.107 &         - &         - &         -  & max AM \\
 & &  &         0.056 &         0.059 &         0.071 &         - &         - &         - & rms AM  \\\hline
\multirow{10}{*}{84720     } & \multirow{5}{*}{A (G8V)}  & \multirow{5}{*}{        4.275} &          0.535 &         0.543 &         0.551 &         1.003 &         1.029 &         1.057 & HZ \\
 & &  &          8.34 &          8.29 &          8.23 &          6.25 &          6.17 &          6.10 & max RV \\
 & &  &          5.73 &          5.69 &          5.65 &          4.19 &          4.13 &          4.08 & rms RV \\
 & &   &         0.240 &         0.244 &         0.248 &         0.461 &         0.474 &         0.487& max AM \\
 & &  &         0.213 &         0.216 &         0.219 &         0.399 &         0.410 &         0.421 & rms AM \\\cline{2-10}
 &  \multirow{5}{*}{B (M0V)}  &  \multirow{5}{*}{        3.364}  &         0.242 &         0.242 &         0.244 &         0.462 &         0.468 &         0.474& HZ \\
 & &  &         15.40 &         15.40 &         15.33 &         11.27 &         11.20 &         11.14 & max RV \\
 & &  &         10.75 &         10.75 &         10.70 &          7.78 &          7.73 &          7.68 & rms RV \\
 & &  &         0.170 &         0.170 &         0.172 &         0.329 &         0.333 &         0.337  & max AM \\
 & &  &         0.153 &         0.153 &         0.154 &         0.292 &         0.296 &         0.300 & rms AM  \\\hline
\multirow{2}{*}{87895     } & A (G2V) &         0.371  & - & - & - & - & - & -  & HZ\\\cline{2-10}
 &  B (K2V) &         0.312  & - & - & - & - & - & -  & HZ  \\\hline
\end{tabular}
\end{minipage}
\end{table*}

\begin{table*}
 \centering
 \begin{minipage}{160mm}
  \caption{Continuation of Table \ref{tab2}. Radial velocity (RV) amplitudes are given in [cm/s], astrometric (AM) amplitudes in [$\mu as$]. 
  The critical planetary semimajor axis $a_{crit}$ as well as the HZ borders are given in [AU].\label{tab4}}
   \begin{tabular}{@{}c|c|c|ccccccl@{}}\hline
HIP ID & comp. & $a_{crit}$ &  inner AHZ &  inner EHZ &  inner PHZ  & outer PHZ  & outer EHZ   & outer AHZ & \\ \hline 
\multirow{10}{*}{93825     } & \multirow{5}{*}{A (F8V)}  & \multirow{5}{*}{        5.623} &          1.289 &         1.311 &         1.337 &         2.421 &         2.505 &         2.581 & HZ \\
 & &  &          3.71 &          3.68 &          3.65 &          2.79 &          2.74 &          2.71 & max RV \\
 & &  &          2.54 &          2.51 &          2.49 &          1.85 &          1.82 &          1.79 & rms RV \\
 & &   &         0.185 &         0.189 &         0.193 &         0.358 &         0.371 &         0.383& max AM \\
 & &  &         0.168 &         0.170 &         0.174 &         0.315 &         0.326 &         0.336 & rms AM \\\cline{2-10}
 &  \multirow{5}{*}{B (F8V)}  &  \multirow{5}{*}{        5.575}  &         1.254 &         1.274 &         1.300 &         2.358 &         2.438 &         2.512& HZ \\
 & &  &          3.79 &          3.76 &          3.72 &          2.84 &          2.80 &          2.76 & max RV \\
 & &  &          2.59 &          2.57 &          2.54 &          1.89 &          1.86 &          1.83 & rms RV \\
 & &  &         0.183 &         0.186 &         0.190 &         0.353 &         0.365 &         0.377  & max AM \\
 & &  &         0.165 &         0.168 &         0.171 &         0.311 &         0.321 &         0.331 & rms AM  \\\hline
\multirow{2}{*}{101916    } & A (G1IV) &         0.754  & - & - & - & - & - & -  & HZ\\\cline{2-10}
 &  \multirow{5}{*}{B (K2IV)}  &  \multirow{5}{*}{        0.381}  &         0.155 &         0.159 &         0.233 &         0.261 &         0.381 &         0.381& HZ \\
 & &  &         38.16 &         37.73 &         31.89 &         30.40 &          - &          - & max RV \\
 & &  &         25.63 &         25.31 &         20.91 &         19.75 &          - &          - & rms RV \\
 & &  &         0.045 &         0.046 &         0.069 &         0.078 &         - &         -  & max AM \\
 & &  &         0.031 &         0.032 &         0.047 &         0.053 &         - &         - & rms AM  \\\hline
\multirow{10}{*}{106972    } & \multirow{5}{*}{A (M2)}  & \multirow{5}{*}{        1.034} &          0.322 &         0.330 &         0.338 &         0.588 &         0.616 &         0.640 & HZ \\
 & &  &         20.73 &         20.50 &         20.28 &         15.94 &         15.65 &         15.41 & max RV \\
 & &  &         13.99 &         13.82 &         13.65 &         10.35 &         10.12 &          9.93 & rms RV \\
 & &   &         0.074 &         0.076 &         0.077 &         0.139 &         0.147 &         0.153& max AM \\
 & &  &         0.053 &         0.054 &         0.055 &         0.096 &         0.101 &         0.105 & rms AM \\\cline{2-10}
 &  \multirow{5}{*}{B (M4)}  &  \multirow{5}{*}{        0.865}  &         0.179 &         0.183 &         0.185 &         0.339 &         0.351 &         0.359& HZ \\
 & &  &         31.41 &         31.08 &         30.92 &         23.35 &         22.99 &         22.76 & max RV \\
 & &  &         21.63 &         21.40 &         21.28 &         15.73 &         15.46 &         15.28 & rms RV \\
 & &  &         0.053 &         0.055 &         0.055 &         0.103 &         0.107 &         0.110  & max AM \\
 & &  &         0.039 &         0.040 &         0.040 &         0.074 &         0.077 &         0.078 & rms AM  \\\hline
\multirow{10}{*}{114922    } & \multirow{5}{*}{A (M1)}  & \multirow{5}{*}{        0.897} &          0.239 &         0.245 &         0.251 &         0.435 &         0.455 &         0.473 & HZ \\
 & &  &         24.59 &         24.31 &         24.05 &         18.89 &         18.53 &         18.24 & max RV \\
 & &  &         16.60 &         16.40 &         16.20 &         12.30 &         12.03 &         11.80 & rms RV \\
 & &   &         0.050 &         0.051 &         0.053 &         0.094 &         0.099 &         0.103& max AM \\
 & &  &         0.037 &         0.038 &         0.039 &         0.068 &         0.071 &         0.074 & rms AM \\\cline{2-10}
 &  \multirow{5}{*}{B (M2)}  &  \multirow{5}{*}{        0.924}  &         0.266 &         0.272 &         0.282 &         0.480 &         0.504 &         0.528& HZ \\
 & &  &         22.83 &         22.60 &         22.23 &         17.67 &         17.32 &         17.00 & max RV \\
 & &  &         15.33 &         15.16 &         14.89 &         11.41 &         11.14 &         10.88 & rms RV \\
 & &  &         0.053 &         0.055 &         0.057 &         0.100 &         0.105 &         0.111  & max AM \\
 & &  &         0.039 &         0.040 &         0.042 &         0.071 &         0.075 &         0.078 & rms AM  \\\hline
\multirow{10}{*}{116132    } & \multirow{5}{*}{A (M4)}  & \multirow{5}{*}{       10.619} &          0.158 &         0.158 &         0.158 &         0.310 &         0.310 &         0.312 & HZ \\
 & &  &         30.77 &         30.77 &         30.77 &         22.02 &         22.02 &         21.95 & max RV \\
 & &  &         21.72 &         21.72 &         21.72 &         15.51 &         15.51 &         15.46 & rms RV \\
 & &   &         0.204 &         0.204 &         0.204 &         0.401 &         0.401 &         0.404& max AM \\
 & &  &         0.165 &         0.165 &         0.165 &         0.323 &         0.323 &         0.325 & rms AM \\\cline{2-10}
 &  \multirow{5}{*}{B (M5)}  &  \multirow{5}{*}{        7.091}  &         0.076 &         0.076 &         0.076 &         0.152 &         0.152 &         0.152& HZ \\
 & &  &         60.97 &         60.97 &         60.97 &         43.14 &         43.14 &         43.14 & max RV \\
 & &  &         43.07 &         43.07 &         43.07 &         30.45 &         30.45 &         30.45 & rms RV \\
 & &  &         0.184 &         0.184 &         0.184 &         0.369 &         0.369 &         0.369  & max AM \\
 & &  &         0.149 &         0.149 &         0.149 &         0.298 &         0.298 &         0.298 & rms AM  \\\hline
\end{tabular}
\end{minipage}
\end{table*}

\begin{table*}
 \centering
 \begin{minipage}{80mm}
  \caption{Transit depths ($TD$), visual brightness ($V$, WDS) and planetary period ($P_p$) ranges are given for potentially transiting planets in the HZs of 
those selected binary systems with $I\approx 90^\circ$.\label{tab5}}
  \begin{tabular}{@{}c|c|ccc@{}}
 \hline
		HIP ID  & comp. & $V$ [mag] & $TD$ [ppm] & $P_{p}$ [D]  \\\hline
\multirow{2}{*}{14669}  &   A   &  10.32  	&  128   &    270.80  - 380.35 \\
		        & B    &  12.5	& 369   &    235.67  - 331.14   \\\cline{1-5}
\multirow{2}{*}{31711}  &   A  & 6.32	&    78 &      338.38  - 476.35  \\
			&  B    &  9.84	&  187  &     270.91  - 380.10   \\\cline{1-5}
\multirow{2}{*}{64241}  &   A   &  4.85 	& 44 &      372.23           \\
		      &   B   &   5.53	& 79  &     346.09  -  382.82       \\\cline{1-5}
\multirow{2}{*}{64797}  &   A  &   6.66	& 171  &     294.45  -  412.52  \\
			& B   & 	9.5 &  198   &    260.18  -  364.81\\\hline
\end{tabular}
\end{minipage}
\end{table*}

\section{Tidal Locking}\label{sec:tides}
An orbital state, where the planet rotates around its own axis with the same speed as it orbits its host-star - much like the Moon around the Earth -
is called 1:1 spin-orbit resonance.
A star-planet system might evolve into such a state due to tidal interactions (see e.g. \citet{murray-dermott-1999}). Therefore, this state is often referred to as tidal lock. 
Since a tidal locking potentially adds additional instabilities to a planet's climate \citep{kite-et-al-2011}, regions where 1:1 spin orbit resonances
occur are usually excluded from HZs. 
\citet{kasting-et-al-1993} used an equation dating back to \citet{peale-1977} to calculate the distance up to which a planet would be 
tidally locked in a time-span equal to age of the Solar System. Inserting such values as chosen in \citet{kasting-et-al-1993} the 
simple estimate reads:
\begin{equation}
r_{TL}\approx 0.46 \left(\frac{AU}{M_\odot^{1/3}}\right)\; m_\star^{1/3}
\end{equation}
with $r_{TL}$ denoting the tidal locking radius in AU and $m_\star$ the mass of the host star in $M_\odot$.
Applying this estimate to our selected systems indicates that all HZs in M-M binaries fall at least partly in the Tidal Locking Zone.
However, tidal evolution of a planet in a binary system is much more involved than simple two body dynamics can account for, 
as the angular momentum transfer between 
the host-star-planet system and the secondary needs to be included in the model.
\citet{eggleton-2006} provides analytical estimates for the tidal evolution of stellar hierarchical triple systems showing that in fact
many possible resonant states other than 1:1 spin-orbit locking exists for the inner pair although with different degrees of stability. 
\citet{wu-murray-2003} and \citet{fabrycky-tremaine-2007} investigated the possibility for tidal migration of planets due to mutually inclined massive perturbers via Kozai cycles
\citep{kozai-1962}. Yet, as point out by \citet{correia-et-al-2011}, only quadrupolar secular expansions had been used to evaluate the planet's eccentricity, 
which give inaccurate results for low inclination configurations such as discussed in the study at hand \citep{lee-peale-2003}. 
Similar to \citet{eggleton-2006}, \citet{correia-et-al-2011} show that tidal interactions in inclined hierarchical triple systems can produce many different outcomes, 
especially when the component's changes in obliquity are taken into account.
Their system's final states included transformations of retrograde to prograde motion and vice versa, 
a decay of mutual inclination and rapid circularization of the inner planetary orbit as well as tidally induced migration.
As more detailed tidal interaction models require knowledge of the stellar radii  \citep{correia-et-al-2011,eggleton-2006}, 
the model dependence of radii for M-dwarfs adds another source of uncertainty, see e.g. \citet{muirhead-et-al-2012}.

The lack of accurate analytical tools to study the influence of tidal interactions in planar S-Type configurations 
as well as the wealth of possible final states depending on the system's initial conditions put a detailed analysis of 
the planet-binary system's tidal evolution beyond the scope of this work.

\section{Summary}\label{sec:sum}
Applying the analytic methods presented in \citep{eggl-et-al-2012b,eggl-et-al-2012}
we have shown that 17  out of 19 binary star systems with well determined stellar and orbital parameters 
close to the Solar System allow for dynamically stable Earth-like planets in circumstellar habitable zones (HZs).
Four of these habitable systems feature F, three feature G, six feature K, and nine feature M class stars. 
Not surprisingly, M-M binary constellations offer the best chances for detecting planets in HZs via radial velocity observations.
However, determining habitability in M star doublets may require additional considerations such as 
tolerable stellar X-ray and EUV fluxes \citep{lammer-et-al-2011} or the system's potential for tidally locking the planet to its host star (see section \ref{sec:tides}).
Habitable planets in systems featuring G-type stars have RV amplitudes comparable to the ones found for 
$\alpha$ Centauri AB \citep{eggl-et-al-2012b, guedes-et-al-2008}. The systems HIP 14699, 30920, 106972, 114922 or 116132 
would be promising candidates to search for terrestrial planets in their HZs, as they offer best case RV semi-amplitudes comparable 
to $\alpha$ Centauri B b \citep{dumusque-et-al-2012}. 
Four of the 17 systems would allow for transiting planets in HZs, which could be detected using current technology. Their mid transit depths
were estimated to lie between 44 and 369 ppm with planetary periods ranging from 235 to 476 days.   
Astrometric signal amplitudes for Earth-like planets in all the investigated systems' HZs are, in contrast, well below \mbox{1 $\mu as$}. 
Therefore, dedicated missions such as NEAT \citep{neat-2011} will be required in order to detect habitable worlds in binary stars via astrometry.    
A sample of 19 systems does not offer the possibility to construct a reasonable statistical analysis 
on the number of potentially habitable binary star systems in the Solar neighborhood.
More precise data on spectral types and orbital elements of nearby double stars are required in this respect.
Nevertheless, our findings indicate that including binary star systems with $1<a_b<100$ in observational campaigns has the potential to enhance our chances of finding habitable worlds.
  
\section*{Acknowledgments}
The authors would like to acknowledge the support of FWF projects S11608-N16 (EP-L \& SE), P20216-N16 (SE, EP-L \& BF) and P22603-N16 (EP-L \& BF). 
SE and EP-L would like to thank the Institute for Astronomy and NASA Astrobiology Institute
at the University of Hawaii for their hospitality during their visit when some of the ideas for this work were developed.
NH acknowledges support from the NASA As-
trobiology Institute under Cooperative Agreement
NNA09DA77A at the Institute for Astronomy,
University of Hawaii, and NASA EXOB grant NNX09AN05G.
SE acknowledges the support of 
University of Vienna's Forschungsstipendium 2012.

%\bibliographystyle{mn2e}
%\bibliography{barbara}

\onecolumn

\appendix

\section{maximum and rms signal amplitudes}\label{ap:amp}
Following \citet{beauge-et-al-2007} \& \citet{eggl-et-al-2012b} the RV amplitude a planet causes on its host star is given by
 \begin{equation}
RV=\frac{\sqrt{G} m_1\sin{I}}{\sqrt{m_0 + m_1}}\frac{ e \cos{\omega} + \cos(f + \omega) }{\sqrt{a (1- e^2) }},
\end{equation} 
where $G$ denotes the gravitational constant, $m_0$ and $m_1$ are the host-star's and planet's masses. 
The quantities $a$, $e$, $I$ and $\omega$ denote the planet's semimajor axis and eccenticity, the system's inclination to the plane of the sky and the planet's 
argument of pericenter respectively. The planet's true anomaly is represented by $f$. 
We can write the maximum possible radial velocity (RV) amplitude caused by a terrestrial planet in a
circumstellar orbit around one binary component as follows:
\begin{equation}
max\; RV = \frac{\sqrt{\mathcal{G}}\;m_1 \sin{I}}{\sqrt{m_0 + m_1}}\frac{(1+e^{max})}{\sqrt{a (1 - (e^{max})^2)}}.
\end{equation} 
The maximum possible eccentricity the planet can acquire due to gravitational interaction with the double star is denoted by $e^{max}$ \citep{eggl-et-al-2012}.
Expressions for the root mean square (rms) values of the RV signal are given as follows \citep{eggl-et-al-2012b}:
\begin{equation}
rms\; RV  =  \frac{1}{2\pi}\left[\iint^{2\pi}_0 RV^2 dM d\omega \right]^{1/2}= \frac{\sqrt{\mathcal{G}}\;m_1|\sin{I}|}{\sqrt{2 a (m_0 + m_1)}}
\end{equation}
In a similar manner we can use the formalism applied in \citet{pourbaix-2002} to determine maximum astrometric (AM) signal strengths:
 \begin{equation}
max\; AM = \frac{\mu a (1 + e^{max})}{d}
\end{equation}  
where $d$ is the observers distance to the observed system, and $\mu=m_1/(m_0+m_1)$ is the planet-host-star system's mass ratio. 
The astrometric rms amplitudes are given by: 
\begin{equation}
rms\; AM =\frac{\mu a}{2d}\left[3 + \frac{9}{2} \langle e^2 \rangle  + \left(1 + \frac{3}{2} \langle e^2 \rangle \right) \cos(2 I)\right]^{1/2}\label{eq:rhorms2}
\end{equation} 
Here, $\langle e^2 \rangle$ is the averaged squared planetary eccentricity. An analytic expression for $\langle e^2 \rangle$ can be found in \citet{georgakarakos-2003,georgakarakos-2005}.
For a detailed derivation of all rms and maximum signal amplitudes see \citet{eggl-et-al-2012b}.

\label{lastpage}

\end{document}